\documentclass[aps,showpacsx,twocolumn]{revtex4-2}
\usepackage{float}
\usepackage{graphicx}
\usepackage{caption}
\usepackage{amsmath}
\usepackage{array}
\usepackage{bm}
\usepackage{amssymb}
\usepackage{amsfonts}
\usepackage{amssymb}
\usepackage{color}
\usepackage[colorlinks, linkcolor=red,anchorcolor=green,citecolor=blue]{hyperref}

\begin{document}
\title{Casimir effect in kinetic theory}
\author{Xingyu Guo$^{1}$}
\author{Jiaxing Zhao$^{2}$}
\author{Pengfei Zhuang$^{2}$}
\affiliation{$^1$Guangdong Provincial Key Laboratory of Nuclear Science, Institute of Quantum Matter, South China Normal University, Guangzhou 510006, China\\
$^{2}$ Physics Department, Tsinghua University, Beijing 100084, China}

\begin{abstract}
We study Casimir effect in equilibrium and non-equilibrium photon gas in the frame of quantum kinetic theory for $U(1)$ gauge field. We derive first the transport, constraint and gauge fixing equations for the photon number distribution from Maxwell's equations, and then calculate the energy variation and Casimir force for a finite system by considering boundary condition on the surface of the system. The Casimir force in vacuum is suppressed by the thermal motion of photons in equilibrium state, when considering two adiabatic plates. In non-equilibrium state, the photon induced Casimir force oscillates and decays with time and finally disappears.  
\end{abstract}
\maketitle

The zero-point energy in vacuum state is an important quantum effect. It is infinite for electromagnetic field because of the existence of infinitely many field oscillators. This infinite contribution is normally considered as a background of the physical problems in excited states and therefore plays no role in most of the cases. However, it is necessary to take into account the zero-point energy when it changes due to the change in the volume or the boundary condition of the physical system. The Casimir effect is such a typical example. The Casimir effect deals with the variation of the vacuum energy of the electromagnetic field inside a cavity limited by a perfectly conducting surface. In the case of two parallel plates the vacuum energy increases by increasing the distance $a$ between the two plates and leads to an attractive force acting on the plate per unit area~\cite{Casimir:1948dh},
\begin{equation}
\label{casimir}
F_0 = -{\pi^2\over 240 a^4}.
\end{equation}       
The Casimir force, firstly observed by Lamoreaux~\cite{PhysRevLett.78.5}, depends strongly on the material properties~\cite{refId0,Pirozhenko:2008tq,Lambrecht:1999vd}, surface properties~\cite{Klimchitskaya:2000zz,Lambrecht:2008eux} and the shape of the plates~\cite{Marachevsky:2007da,BLOCKI1977427}. Important and recent advances of the Casimir effect include repulsive Casimir force~\cite{Munday:2009fgb}, lateral Casimir force~\cite{PhysRevLett.118.133605,PhysRevLett.121.130401}, Casimir torque~\cite{Somers2018} and tailoring of the Casimir force~\cite{Intravaia2013}. 
 
The Casimir effect in medium with nonzero photon number is very different from that in vacuum. For a photon gas in equilibrium state, the total Casimir force comes from both the thermal and quantum fluctuations of the electromagnetic field~\cite{Bostrom:2000zza,Brevik:2006jw}. The thermal Casimir force which should dominate at large separation is widely discussed, for instance, in thermal quantum field theory, Lifshitz formalism, stress-tensor approach and multiple-scattering formulation~\cite{Geyer:2008wb,PhysRevLett.104.040403,WorldScientific2005}. Experimentally, the thermal Casimir force is observed in 2011~\cite{Sushkov:2010cv}. For a photon gas in non-equilibrium state, the Casimir effect is investigated in different systems~\cite{JOULAIN200559}, such as parallel plates~\cite{Antezza:2007lpu,PhysRevLett.97.223203}, sphere plates~\cite{Kruger:2011zd} and plane-sphere setups~\cite{Canaguier-Durand:2010hwu}. The thermal radiation and heat-flux between two parallel plates in a non-equilibrium system is also discussed in detail~\cite{PhysRevB.4.3303,PhysRevLett.106.210404}.  
 
For the Casimir force in medium which is controlled by the interaction potential, ${\bm F}=-{\bm \nabla}V({\bm r})$, a natural question is the definition of the potential $V$. For a photon gas in equilibrium state, if we do not consider the heat exchange between the plate and the photon gas, the potential is just the internal energy ${\mathcal E}$, while if there is enough time for the plate to exchange heat with the photon system, the free energy ${\mathcal F}$ is taken as the potential. Considering the fact that metal plates are usually easy to exchange heat with the photon gas, free energy ${\mathcal F}$ is used in the previous studies~\cite{Geyer:2008wb,PhysRevLett.104.040403,WorldScientific2005,Sushkov:2010cv}. However, if the plates are adiabatic, the internal energy ${\mathcal E}$ should be taken as the potential. On the other hand, ${\mathcal F}$ is a state function and can only be applied to equilibrium states, but the internal energy ${\mathcal E}$ can still be well-defined in non-equilibrium systems through the particle energy $\epsilon$ and distribution function $f$, ${\mathcal E}=\int dx dp \epsilon(p)f(x,p)$.  
  
An often used approach to describe the evolution of a non-equilibrium system in phase space is the quantum kinetic theory in the frame of Wigner function formalism, by taking Wigner transformation of non-relativistic Schr\"odinger or relativistic Dirac equation for systems of quantum mechanics~\cite{Elze:1986qd, zhuang_relativistic_1996} and Dyson-Schwinger equation for systems of quantum field~\cite{yang_effective_2020, wang_local_2020}. A group of kinetic equations for polarized photons with collision terms is recently derived~\cite{hattori_wigner_2021} and used to study the chiral vortical effect of photons~\cite{huang_zilch_2020} by taking semi-classical expansion, similar to the treatment of photon transport phenomena based on the Berry phase\cite{bliokh_topological_2004,yamamoto_photonic_2017,huang_chiral_2018}. Considering the massless limit of quantum kinetic equations for fermions~\cite{chen_berry_2013,chen_collisions_2015,hidaka_relativistic_2017,huang_complete_2018,sun_chiral_2018,liu_chiral_2019,weickgenannt_kinetic_2019,gao_relativistic_2019,yang_effective_2020}, it is found that the chiral fermions share similar properties with gauge bosons, such as the Berry phase~\cite{son_berry_2012,mueller_chiral_2018}, the side-jump effect~\cite{hidaka_relativistic_2017,yang_side-jump_2018,abbasi_magneto-transport_2018,carignano_consistent_2018} and the chiral vortical effect~\cite{gao_chiral_2019,hidaka_non-equilibrium_2019,morales-tejera_out_2020,prokhorov_chiral_2020}. The coupled kinetic equations for gauge field and chiral fermion field are recently applied to the study of high-energy nuclear collisions with gluons and quarks and of Dirac and Weyl semi-metals with photons and electrons~\cite{son_chiral_2013,gorbar_anomalous_2018,dantas_magnetotransport_2018}. 
  
While most of the applications of the quantum kinetic theory for gauge bosons and chiral fermions are through the semi-classical approximation, one needs to solve the full kinetic equations for those phenomena which do not have classical correspondence, for instance the pair production~\cite{PhysRev.82.664, zhuang_relativistic_1996} in strong electromagnetic field and the Casimir effect discussed here. The other characteristic to discuss Casimir effect in the frame of kinetic theory is the boundary condition of the kinetic equations, similar to the treatment of bound state problems in quantum mechanics. In this paper we are interested in the photon Wigner function which satisfies the quantum kinetic equations and disappears on the boundary surface.  
 
We study the Casimir effect in medium in the frame of quantum kinetic theory for electromagnetic field. We will focus on the photon number contribution to the Casimir force in equilibrium and non-equilibrium states, by taking the internal energy ${\mathcal E}$ as the interaction potential for the calculation of the Casimir force. The paper is organized as follows. We first derive and solve the transport and constraint equations for the photon Wigner function in the cavity constructed by two plates. From the energy-moment tensor in terms of the photon distribution function, we then calculate the energy density of the system and the Casimir force acting on the plates in equilibrium and non-equilibrium states. We will also discuss the relation between the internal energy ${\mathcal E}$ and free energy ${\mathcal F}$ and the difference between equilibrium and non-equilibrium systems.      
 
The Wigner function for electromagnetic field $A_\mu(x)$ is defined as
\begin{equation}
\label{wigner}
G_{\mu\nu}(x,p) = \int d^4 y e^{-ip\cdot y}\langle A_{\mu}(x+y/2)A_\nu(x-y/2)\rangle,
\end{equation}
where $\langle\cdots\rangle$ means the ensemble average of the field product. Since we are interested in the full solution of the Wigner function, we will not explicitly express the $\hbar$-dependence here and in the following. Note that different from the fermion Wigner function~\cite{Elze:1986qd} the photon Wigner function is hermitian,  
\begin{equation}
\label{hermitian}
G^*_{\nu\mu} = G_{\mu\nu}.
\end{equation}	
 
From the Lagrangian density $\mathcal{L}=-1/4F_{\mu\nu}F^{\mu\nu}$ for a system of photon gas with gauge field tensor $F_{\mu\nu}=\partial_\mu A_\nu-\partial_\nu A_\mu$, the Wigner transformation of the field equation
\begin{equation}
\label{field}
\partial_\nu\partial^\nu A_\mu-\partial_\mu\partial^\nu A_\nu =0
\end{equation}
leads to the kinetic equation for the photon Wigner function,
\begin{equation}
\label{kinetic}
\bar p_\sigma\bar p^\sigma G_{\mu\nu}-\bar p_\mu \bar p^\sigma G_{\sigma\nu}=0
\end{equation}
with the modified momentum operator $\bar p_\mu = p_\mu+i\partial_\mu/2$. For a system of quantum electrodynamics (QED), there will be collision terms on the right hand side of the kinetic equation due to the interaction among photons and fermions, and the photon Wigner function is coupled to the fermion Wigner function.
 
For simplicity we choose the Lorenz gauge $\partial_\mu A^\mu=0$, the kinetic equation is then reduced to
\begin{equation}
\label{kinetic1}
\bar p_\sigma \bar p^\sigma G_{\mu\nu}=0,
\end{equation}
and the gauge fixing in phase space becomes 
\begin{equation}
\label{kinetic2}
\bar p^\mu G_{\mu\nu} =0.
\end{equation}
Separating the real and imaginary parts of the two equations and making use of the hermitian condition (\ref{hermitian}), we obtain the covariant transport and constraint equations for the photon Wigner function,
\begin{eqnarray}
&& p^\mu\partial_\mu G_{\mu\nu} ^\pm = 0,\label{transport}\nonumber\\
&& (p^2-\partial^2/4)G_{\mu\nu}^\pm = 0,\label{offshell}\nonumber\\
&& p^\mu G_{\mu\nu}^\pm +i\partial^\mu/2 G_{\mu\nu}^\mp = 0
\label{gauge}
\end{eqnarray}
with the definition of $G_{\mu\nu}^\pm =(G_{\mu\nu}\pm G_{\nu\mu})/2$. From the hermitian condition $G_{\mu\nu}^-$ is purely imaginary and $G_{\mu\nu}^+$ is purely real. In classical kinetic theory, the particle distribution function $f$ satisfies the Boltzmann equation and the on-shell condition $(p^2-m^2)f(x,p)=0$~\cite{harris2004introduction}. In quantum case the Boltzmann equation and on-shell condition are replaced by their quantum analogies, namely the quantum transport equation and the off-shell constraint~\cite{zhuang_relativistic_1996,huang_complete_2018}. For gauge bosons, the above three kinetic equations characterize respectively the transport properties, the off-shell effect, and the gauge fixing condition for the photon Wigner function.  
 
We can express the phase-space version $j_\mu(x,p)$ of the probability current $J_\mu(x) = F_{\mu\nu}(x)A^\nu(x)=\int d^4p j_\mu(x,p)$ for the $U(1)$ gauge field in terms of the Wigner function. A direct Fourier transformation leads to 
\begin{equation}
\label{current1}
j_\mu = -i(\bar p_\mu G_\nu^{~\nu}-\bar p_\nu G_\mu^{~\nu}).
\end{equation}
Arising from the Klein-Gordon equation $\partial^2 A_\mu=0$ for the gauge field under the Lorenz gauge, there is the problem of negative probability in coordinate space and complex current in phase space~\cite{}. We then take its real part
\begin{equation}
\label{current2}
\tilde j_\mu =\partial_\mu/2 G_\nu^{~\nu}-\partial^\nu G_{\mu\nu}^+,
\end{equation}
where we have used the gauge fixing condition in (\ref{gauge}). Note that for gauge bosons the current is not conserved, which is not strange for electromagnetic field since we have $\partial_\mu J^\mu = F_{\mu\nu}F^{\mu\nu}\sim |E|^2-|B|^2$ with $E$ and $B$ as the electromagnetic field strengths. Taking the similar procedure, the phase-space version of the axial current $J_5^\mu = \epsilon^{\mu\nu\sigma\rho}A_\nu F_{\sigma\rho}$ for the gauge field can be written as 
\begin{equation}
\label{current3}
\tilde j_5^\mu = -i\epsilon^{\mu\nu\sigma\rho}p_\nu G_{\sigma\rho}^-.
\end{equation}
 
For the energy-momentum tensor $T_{\mu\nu}=F_{\mu\sigma}F_{~\nu}^\sigma-g_{\mu\nu}F_{\sigma\rho}F^{\rho\sigma}/4$, its phase-space expression which is the starting point of discussing the Casimir effect can be written as  
\begin{eqnarray}
\label{tensor1}
t_{\mu\nu} &=& p_\mu p_\nu G^\sigma_{~\sigma}-\left(p_\mu p^\sigma\right)G_{\nu\sigma}^+\nonumber\\
&&-p_\nu p^\sigma G_{\mu\sigma}^+p^2 G_{\mu\nu}^+\nonumber\\
&&-g_{\mu\nu}/2\left(p^2 G^\sigma_{~\sigma}-p^\sigma p^\rho G_{\rho\sigma}^+\right).
\end{eqnarray}
In deriving the above expression we have used the fact that $T^{\mu\nu}$ can be modified by a full derivative term. It is straightforward to see that the energy and momentum of the system are conserved, 
\begin{equation}
\label{conservation}
\partial_\mu t^{\mu\nu}=0.
\end{equation}
It is interesting to notice that the vector current and energy-momentum tensor depend on $G_{\mu\nu}^+$ while the axial vector current depends on $G_{\mu\nu}^-$. Due to the gauge fixing condition in (\ref{gauge}) the two are related to each other.
 
We now simplify the transport and constraint equations (\ref{gauge}) for the Wigner functions $G_{\mu\nu}^\pm$. For a pure $U(1)$ gauge theory, photons should be always on the mass shell, namely the condition $p^2 G=0$ holds for any Wigner function $G$, and its phase space correlation can be described by a scalar function. Substituting the factorization 
\begin{equation}
\label{factorization}
G_{\mu\nu}^\pm(x,p) = C_{\mu\nu}^\pm(p)f_\pm(x,p)\delta(p^2)
\end{equation}
into the kinetic equations, the tensors $C_{\mu\nu}^\pm(p)$ as functions of momentum satisfy the constraints
\begin{equation}
\label{tensor2}
p^\mu C_{\mu\nu}^\pm =0
\end{equation} 
which lead to a general solution
\begin{eqnarray}
C_{\mu\nu}^+ &=& {p_\mu p_\nu\over (p\cdot u)^2}-{2(p_\mu u_\nu -p_\nu u_\mu)\over p\cdot u}+g_{\mu\nu},\nonumber\\
C_{\mu\nu}^- &=& i\epsilon_{\mu\nu\sigma\rho}{p^\sigma u^\rho\over 2p\cdot u},
\end{eqnarray}
where $u_\mu$ is a constant vector satisfying $u^\mu G_{\mu\nu}=0$. The two scalar functions $f_\pm(x,p)$ are characterized by the transport and constraint equations
\begin{eqnarray}
\label{scalar}
&& p\cdot\partial f_\pm = 0,\nonumber\\
&& \partial^2 f_\pm =0,\nonumber\\
&& \left(p_\mu u\cdot\partial - p\cdot u\partial_\mu\right)f_+=0,\nonumber\\
&& \epsilon_{\mu\nu\sigma\rho}\partial^\nu p^\sigma u^\rho f_- =0.
\end{eqnarray}
We can rewrite the vector and axial-vector currents and energy-momentum tensor in terms of the two scalar distributions, 
\begin{eqnarray}
\label{current_tensor}
&& \tilde j_\mu = \partial_\mu f_+,\nonumber\\
&& \tilde j_5^\mu = p^\mu f_-,\nonumber\\
&& t_{\mu\nu} = 2p_\mu p_\nu f_+.  
\end{eqnarray} 
It is clear that $f_+$ is the photon number distribution and $f_-$ describes the axial symmetry breaking.
 
To study the Casimir effect, we next consider the photon Wigner function for a system that is finite in one dimension, by setting two parallel plates at $z=0$ and $z=a$. Taking into account the boundary condition for the electromagnetic fields ${\bm E}$ and ${\bm B}$, 
\begin{eqnarray}
\label{boundary}
&& E_x|_{z=0}=E_y|_{z=0}=E_x|_{z=a}=E_y|_{z=a}=0,\nonumber\\
&& B_z|_{z=0}=B_z|_{z=a}=0,
\end{eqnarray}
the photon longitudinal momentum $p_z$ becomes discontinuous $p_z=n\pi/a$ with $n=-\infty,\cdots,-1,0,1,\cdots\infty$, and the integration $\int_{-\infty}^\infty dp_z/(2\pi)$ is replaced by a summation $1/(2a)\sum_{n=-\infty}^\infty$.
 
We now consider the Casimir effect in equilibrium medium. The equilibrium distribution function can in principle be determined by detailed balance between the lose and gain terms on the right hand side of the transport equation~\cite{DeGroot:1980dk}. The result is well-known that photons at finite temperature $T$ are in the Bose-Einstein state with distribution $f_+(p)=1/(e^{\epsilon_p/T}-1)$ in the rest frame of the medium, where $\epsilon_p=|{\bm p}|$ is the photon energy. It is clear that the global equilibrium distribution satisfies all the kinetic equations (\ref{scalar}).
 
\begin{figure}[H]
	\centering
	\includegraphics[width=0.4\textwidth]{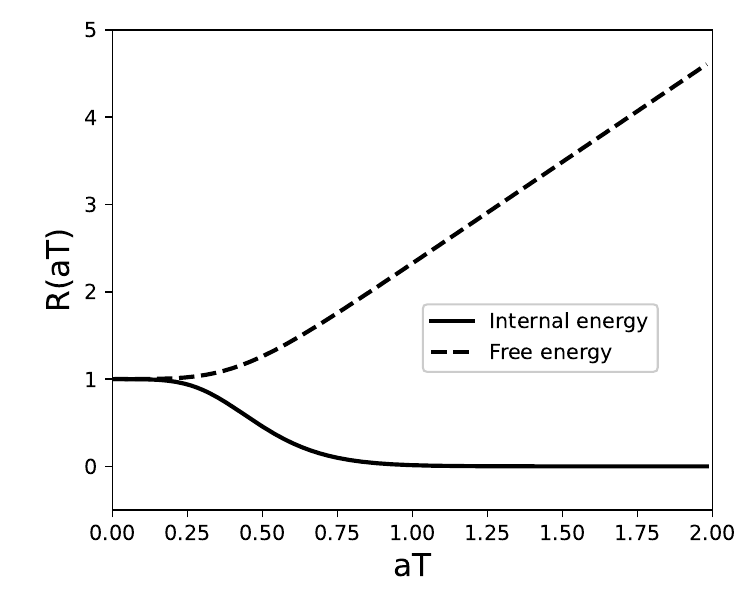}
	\caption{The ratio between the Casimir forces in equilibrium medium and in vacuum as a function of scaled temperature $aT$, calculated by taking internal energy ${\mathcal E}$ (solid line) or free energy ${\mathcal F}$ (dashed line) as the interaction potential.}
	\label{fig1}
\end{figure}

Taking the energy-momentum tensor for the global equilibrium state, the internal energy per unit transverse area of the cavity between the two plates is
\begin{eqnarray}
\label{energy}
{\mathcal E} &=& 2 a \int{d^3{\bm p}\over (2\pi)^3 2\epsilon_p}t_{00}(p)\nonumber\\
&=& 2 a\int{d^3{\bm p}\over (2\pi)^3}\epsilon_pf_+(p),
\end{eqnarray}
where the factor $2$ comes from the two independent spin degrees of freedom of the photons, and the factor $a$ is from the longitudinal integration $\int dz$. The variation of the energy by the boundary condition at the surface can be expressed as  
\begin{equation}
\label{variation}
\Delta {\mathcal E} = 2\left(\sum_{n=0}^\infty -\int_0^\infty dn\right)\int{d^2{\bm p}_\perp\over (2\pi)^2}\epsilon_pf_+(p)
\end{equation}
with photon energy $\epsilon_p=\sqrt{{\bm p}_\perp^2+(n\pi/a)^2}$ and photon transverse momentum ${\bm p}_\perp$ parallel to the two plates, where we have considered the fact that the integrated function is an even function of $n$. Note that, different from the calculation in vacuum~\cite{Casimir:1948dh} where the energy ${\mathcal E}$ with and without considering the boundary condition is divergent but the variation $\Delta {\mathcal E}$ is convergent, here in medium both ${\mathcal E}$ and $\Delta {\mathcal E}$ are convergent due to the suppression from the thermal distribution. The variation induced force per unit area at finite temperature $T$ acting on the plate at $z=a$ is
\begin{equation}
\label{force1}
F_T = -{\partial \Delta {\mathcal E}\over \partial a}.
\end{equation} 
Using the Bose-Einstein distribution, a direct calculation leads to 
\begin{equation}
\label{force2}
F_T = {\pi^2\over 15}T^4-{\pi^2\over a^4}\sum_{n=0}^\infty{n^3\over e^{n\pi/(aT)}-1}.
\end{equation}
Taking into account the background contribution from the vacuum, which can be calculated by replacing the number distribution $f_+$ in (\ref{energy}) and (\ref{variation}) by the zero-point factor $1/2$, the Casimir force in medium becomes $F_m=F_0+F_T$, and the ratio between the forces in medium and in vacuum is 
\begin{eqnarray}
\label{ratio}
R(aT) &=& F_m/F_0\nonumber\\
&=& 1+240\sum_{n=0}^\infty {n^3\over e^{n\pi/(aT)}-1}-16(aT)^4.
\end{eqnarray}
Note that, the ratio does not depend on the distance $a$ and temperature $T$ separately but is a function of the dimensionless parameter $aT$. From the numerical result shown as the solid line in Fig. \ref{fig1}, the ratio drops down continuously with increasing $aT$ and finally disappears. This indicates clearly that, the attractive Casimir force in vacuum is gradually canceled by the thermal motion of photons. 

Our result here seems different from the previous works~\cite{Sushkov:2010cv,Geyer:2008wb,PhysRevLett.104.040403,WorldScientific2005} where the Casimir force is found to be enhanced at finite temperature. The reason for this difference is that we considered different systems. In these previous works the plates are fully thermalized with the photon gas, and therefore the free energy ${\mathcal F}$ is taken to calculate the Casimir force ${\bm F}=-{\bm \nabla}{\mathcal F}$, while in our consideration the plates are assumed to be adiabatic and the Casimir force is calculated from the internal energy ${\bm F}=-{\bm \nabla}{\mathcal E}$. Taking into account the thermodynamic relation between ${\mathcal F}$ and ${\mathcal E}$, ${\mathcal E}={\mathcal F}+\beta\partial {\mathcal F}/\partial\beta\ (\beta=1/T)$, it is easy to check the self-consistency of the two kinds of calculations. The ratio~\cite{Geyer:2008wb}
\begin{equation}
R(aT) = 1-30\sum_{n=1}^\infty {aT\over n\pi}{\sinh({n\pi\over aT})\over \sinh^4({n\pi\over 2aT})}+{16\over 3}(aT)^4
\end{equation}
calculated with ${\mathcal F}$ is shown as the dashed line in Fig.\ref{fig1}. At low temperature with $aT\ll 1$ the ratio behaviors as 
\begin{equation}
R(aT)=1+{16\over 3}(aT)^4,
\end{equation}
and at high temperature with $aT\gg 1$ it becomes a linear increase,
\begin{equation}
R(aT)={60\over \pi^3}\zeta(3)aT.
\end{equation} 

The difference between ${\mathcal F}$ and ${\mathcal E}$ is a general phenomenon when studying an object located in a thermalized system. For instance, in high energy nuclear physics the quarkonium evolution in a quark-gluon plasma depends strongly on the choice of ${\mathcal F}$ or ${\mathcal E}$~\cite{}. Note that the difference disappears automatically in vacuum at $T=0$. The other reason for us to choose internal energy is that, it is more suitable when studying non-equilibrium systems, because the free energy is a state function that describes only equilibrium states.
 
For a non-equilibrium system of photons, the distribution is fully controlled by its initial condition. Suppose the initial distribution is $f_0(t_0,{\bm x}_0,p)$, the kinetic equations (\ref{scalar}) describe the photon free-streaming motion with the solution    
\begin{equation}
\label{none}
f_+(x,p)=f_0(t_0,{\bm x}-{\bm p}/\epsilon_p(t-t_0),p).
\end{equation}
Let's consider a reasonable initial distribution 
\begin{equation}
\label{initial}
f_0(0,{\bm x}_0,p)=e^{-\epsilon_p r_0}
\end{equation}
with initial time $t_0=0$ and initial radius $r_0=|{\bm x}_0|$. Since the photon distribution
\begin{equation}
\label{dis}
f_+(x,p)=e^{-\sqrt{(\epsilon_p{\bm x}-{\bm p}t)^2}}
\end{equation} 
is no longer a constant in coordinate space, the Casimir force acting on the plates depends on the transverse coordinate ${\bm x}_\perp$. We consider the internal energy of the cavity per unit transverse area at coordinate ${\bm x}_\perp=0$ and time $t$,
\begin{eqnarray}
\label{energy2}
{\mathcal E} &=& 2 \int {d^3{\bm p}\over (2\pi)^3}\int dz\epsilon_pf_+(x,p)\nonumber\\
&=& 2 \int {d^3{\bm p}\over (2\pi)^3}\int dz \epsilon_pe^{-\sqrt{(p_\perp t)^2+(\epsilon_pz-p_z t)^2}}.
\end{eqnarray}
Introducing the dimensionless variables $z'=z/a$, $p_z'=ap_z$ and $p'_\perp=ap_\perp$, and considering the boundary condition, the energy variation becomes
\begin{eqnarray}
\label{variation3}
\Delta {\mathcal E} &=&  {1\over a^3}\Delta\bar{\mathcal E},\nonumber\\
\Delta\bar{\mathcal E} &=& {1\over 2\pi}\left(\sum_{-\infty}^\infty-\int_{-\infty}^\infty dn\right)\int_0^\infty dp_\perp'\int_0^1 dz'\nonumber\\
&&\times p_\perp'\epsilon_{p'}e^{-\sqrt{(p'_\perp t/a)^2+(\epsilon_p'z'-n\pi t/a)^2}}
\end{eqnarray}
with $\epsilon_{p'}=a\epsilon_p$. Considering particle's anisotropic motion in non-equilibrium state, the integrated function here is in general not an even function of $p_z\ (n)$, the summation and integration over $n$ starts at $-\infty$.    

Calculating the force per unit area $F_t=-\partial\Delta {\mathcal E}/\partial a$ and taking into account the background contribution from the vacuum, the ratio of the Casimir forces $F_m=F_0+F_t$ in non-equilibrium medium and $F_0$ in vacuum becomes
\begin{eqnarray}
\label{ratio2}
R(t/a) &=& F_m/F_0\nonumber\\
&=& 1-{240\over \pi^2}\left(3\Delta\bar{\mathcal E}(t/a)+{\partial\Delta\bar{\mathcal E}(t/a)\over \partial(t/a)}t/a\right).
\end{eqnarray}
Similar to the calculation in equilibrium state where the ratio depends only on the scaled temperature $aT$, the ratio in non-equilibrium state is controlled only by the scaled time $t/a$. 

\begin{figure}[H]
\centering
\includegraphics[width=0.4\textwidth]{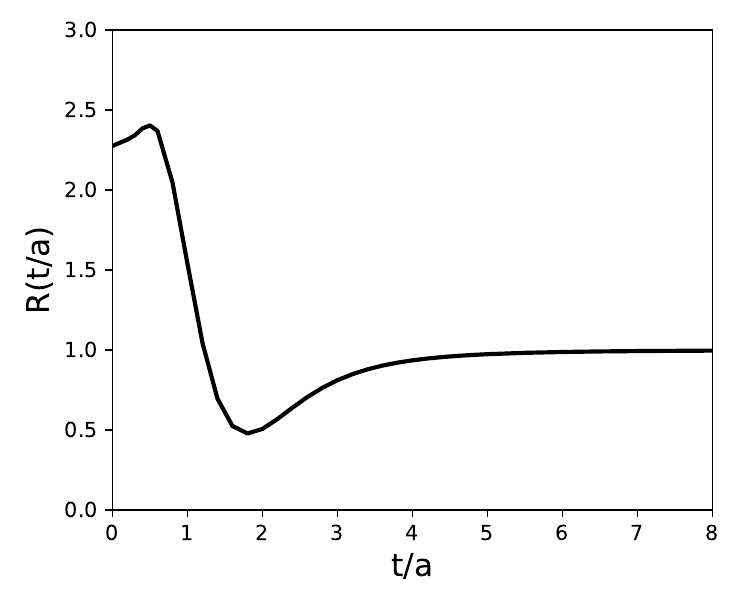}
\caption{The ratio between the Casimir forces in non-equilibrium medium and in vacuum as a function of scaled time $t/a$.}
\label{fig2}
\end{figure}
The numerical result of the ratio as a function of $t/a$ is shown in Fig.\ref{fig2}, its behavior is totally different from that in equilibrium state. Let's first consider the ratio at the starting time $t=0$. While the energy ${\mathcal E}$ is in general convergent for $t>0$, it is divergent at $t=0$ since the momentum integration is divergent at $z=0$. In this case, we can analytically take the 
$z-$integration and then employ the Abel-Plana formula~\cite{hardy1967divergent} to obtain the finite energy variation
\begin{eqnarray}
\label{ev}
\Delta\bar{\mathcal E}(t/a=0) &=& -2\pi\int_0^\infty dx{1\over e^{2\pi x}-1}\big[\pi x^2/4\nonumber\\
&&-xI_1(\pi x)/2-\mathcal L_1(\pi x)\big]\nonumber\\
&=&-0.017
\end{eqnarray}  
and the Casimir ratio
\begin{equation}
\label{r1}
R(t/a=0)=1-{720\over \pi^2}\Delta\bar{\mathcal E}(t/a=0)=2.276,
\end{equation} 
where $I_1$ and $\mathcal L_1$ are modified Bessel function and Struve function. $R\gg 1$ means a strong enhancement of the attractive Casimir force in non-equilibrium medium. With increasing time, the expansion of the medium dilutes the photon gas, and the Casimir force oscillates and decays during the dilution, as a typical behavior in a non-equilibrium transport process. In the limit of $t\to \infty$, the photon distribution $f_+\to 0$ and the contribution to the Casimir force disappears,
\begin{equation}
R(t/a\to\infty)=1.
\end{equation} 

We studied in this paper Casimir effect in medium in the frame of quantum kinetic theory. From Maxwell's equations for electromagnetic field we derived the transport, constraint and gauge fixing equations for the photon number distribution. Considering the boundary condition on the surface of a finite system, we calculated the energy variation and Casimir force in equilibrium and non-equilibrium photon gas. We found that, when the two plates are adiabatic from the photon gas, the attractive Casimir force in vacuum is suppressed in equilibrium medium by the thermal motion of photons and finally vanishes in high temperature limit. In non-equilibrium medium, the force oscillates and decays with time due to the expansion of the medium and the dilution of the photon gas.      
 
Finally we discuss the experimental possibility to measure the non-equilibrium Casimir force. One can place a controllable point-like light source at a fixed position between two parallel plates. When the light source is turned on for a short time, it produces an initial distribution like (\ref{initial}). Then we turn off the light source and measure the pressure acting on a plate. This measured pressure is the combination of the Casimir force and the radiation pressure. However, one can separate these two effects by measuring the pressure's dependence on the distance $a$. The Casimir force has a nontrivial dependence on $a$ as we demonstrated, while the radiation pressure does not depend on $a$. One can imagine other systems to measure the non-equilibrium Casinir force, for instance two plates held at different temperatures. Different systems provide different initial distributions, but the kinetic equations (\ref{scalar}) describing the non-equilibrium evolution of the systems are the same. 
    
\noindent {\bf Acknowledgement}: The work is supported by NSFC grant Nos. 11890712, 12035007 and 12075129, Guangdong Major Project of Basic and Applied Basic Research No. 2020B0301030008, and Science and Technology Program of Guangzhou No. 2019050001.  

\bibliography{transport}
\end{document}